\documentclass[epj-spec]{svjour}
\usepackage{amsmath}
\usepackage{graphicx}
\usepackage{amssymb}
\begin{document}

\title{Quantum state transfer and time-dependent disorder in Quantum Chains}

\author{Daniel Burgarth}
\institute{Institute of Theoretical Computer Science, ETH Zurich,8092 Zurich, Switzerland}
\abstract{
One of the most basic tasks required for Quantum Information Technology
is the ability to connect different components of a Quantum Computer
by \emph{quantum wires} that obey the superposition principle. Since
superpositions can be very sensitive to noise this turns out to be
already quite difficult. Recently, it was suggested to use chains
of permanently coupled spin-1/2 particles (\emph{quantum chains})
for this purpose. They have the advantage that no external control
along the wire is required during the transport of information, which
makes it possible to isolate the wire from sources of noise. We first
give an introduction to basic quantum state transfer and review existing
advanced schemes by other authors. We then show a new result that
demonstrates the stability of the scheme~\cite{Burgarth} against
disorder that is approximately constant during one application of the
channel, but time-dependent with respect to multiple applications.
}

\maketitle
\section{Introduction}

\subsection{Spin chains for quantum state transfer}

It is perhaps surprising that more than twenty years after the theoretical
birth of Quantum Computers only very small prototypes have been built.
One of the main problems is the {}``programming'', i.e. the design of a specific
(time-dependent) Hamiltonian, usually described as a set of discrete
unitary gates. This turns out to be extremely difficult because we
need to connect microscopic objects (those behaving quantum mechanically)
with macroscopic devices that \emph{control} the microscopic behavior.
Even if one manages to find a link between the micro- and the macroscopic
world, such as laser pulses and electric or magnetic fields, 
the connection often introduces not only control but also noise (dissipation
and decoherence~\cite{OPENQUANTUM}) to the microscopic system, and its quantum behavior
is diminished.

\begin{figure}[htbp]
\begin{centering}\includegraphics[width=0.7\columnwidth]{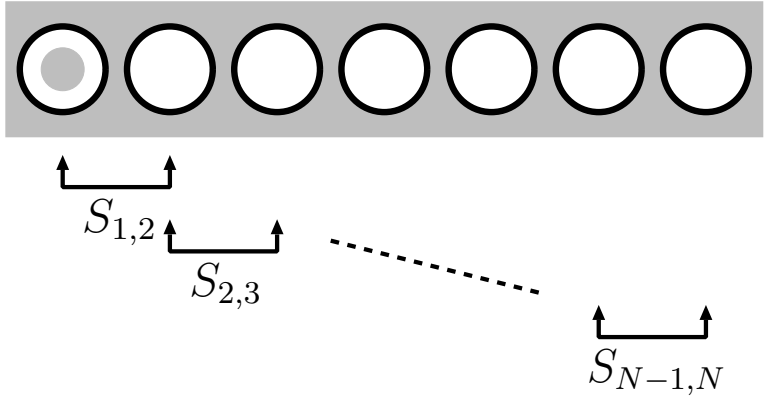}\par\end{centering}

\caption{\label{fig:swapping0}In areas of universal control, quantum states
can easily be transferred by sequences of unitary swap gates $S_{j,k}$
between nearest neighbours.}
\end{figure}

As part of the vision to develop theoretical methods narrowing the
gap between what is imagined theoretically and what can be done experimentally,
S. Bose suggested to use chains of \emph{permanently coupled} quantum
systems for the specific task of quantum communication~\cite{Bose2003}.
Due to the {}``always on'' coupling, these devices can in principle be built
in such a way that they do not require external control to perform
their tasks, just like a mechanical clockwork. This also overcomes
the problem of decoherence as they can be separated from any source
of noise. Although in theory,
the \emph{universal set of gates} on the quantum
computer can be used to transfer quantum states by applying sequences
of two-qubit swap gates (Fig.~\ref{fig:swapping0}), in practice
it is crucial to minimize the required number of quantum gates, as
each gate typically introduces \emph{errors}.%
 In this light it appears costly to perform $N-1$ swap gates between
nearest neighbors to just move a qubit state over a distance of $N$
sites. For example, Shor's algorithm on $N$ qubits can be implemented
by only $\log N$ quantum gating operations~\cite{Cleve2000} if
long distant qubit gates are available. These long distant gates could
consist of local gates followed by a quantum state transfer. If however
the quantum state transfer is implemented as a sequence of local gates,
then the number of operations blows up to the order of $N$ gates.
The quantum state transfer can even be thought of as the source of
the \emph{power of quantum computation}, as any quantum circuit with
$\log N$ gates and \emph{local gates only} can be efficiently simulated
on a classical computer~\cite{Josza2006,Short2006}.

A second reason to consider devices for quantum state transfer is
related to \emph{scalability}. While small quantum computers have
already been built~\cite{Haffner2005}, it is very difficult to build
large arrays of fully controllable qubits. A \emph{black box} that
transports unknown quantum states could be used to build larger quantum
computers out of small components by connecting them. Likewise, quantum
state transfer can be used to connect \emph{different} components
of a quantum computer, such as the processor and the memory. On larger
distances, flying qubits such as photons, ballistic electrons and
guided atoms/ions are considered for this purpose~\cite{Skinner2003,Kielpinski2002}.
However, converting back and forth between stationary qubits and mobile
carriers of quantum information and interfacing between different
physical implementations of qubits is very difficult and not worthy
only for short communication distances. This is the typical situation
one has to face in solid state systems, where quantum information
is usually contained in the states of \emph{fixed objects} such as
quantum dots or Josephson junctions.
In many cases, permanent couplings (and hence quantum chains) are easy to build
in solid state devices. The qubits can be of the \emph{same type}
as the other qubits in the device, so no interfacing is required.
\begin{figure}[htbp]
\begin{centering}\includegraphics[width=0.7\columnwidth]{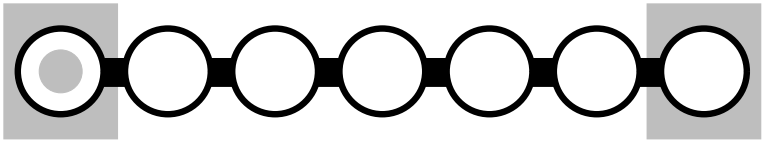}\par\end{centering}

\caption{\label{fig:qchain}Permanently coupled quantum chains can transfer
quantum states without control along the line. Note that the ends
still need to be controllable to initialize and read out quantum states.}
\end{figure}

Finally, an important reason to study quantum state transfer in quantum
chains stems from a more fundamental point of view. Such systems in
principle allow tests of Bell-inequalities and non-locality in solid-state
experiments well before the realization of a full quantum computer. Although
quantum transport is quite an established field, the quantum information
point of view offers many new perspectives. Here, one looks at the
transport of information rather than excitations, and at entanglement~\cite{Lakshminarayan2006,Subrahmanyam2004,Eisert2004,Palma2004}
rather than correlation functions. It has recently been shown that
this sheds new light on well-known physical phenomena such as quantum
phase transitions~\cite{Plenio,Plenio2006,Verstraete,Verstraete2004},
quantum chaos~\cite{Monteiro2006,Twamley2005,Santos2004,Boness2006}
and localization~\cite{Winter,Apollaro2006,Burrell2007}. Furthermore,
quantum information takes on a more \emph{active} attitude. The correlations
of the system are not just calculated, but one also looks at how they
may be \emph{changed}.

\subsection{Physical implementations and experiments\label{sec:Implementations-and-experiments}}

\begin{figure}
\begin{centering}\includegraphics[width=0.7\columnwidth]{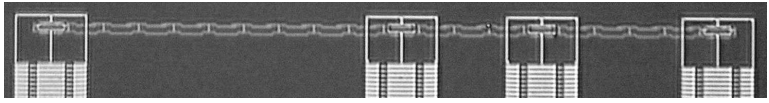}\par\end{centering}

\caption{\label{fig:A-quantum-chain}A quantum chain consisting of $N=20$
flux qubits~\cite{Bruder2005} (picture and experiment by Floor Paauw,
TU Delft). The chain is connected to four larger SQUIDS for readout
and gating.}
\end{figure}

As we have seen above, the main advantage of state transfer with quantum
chains is that the qubits can be of the same type as those used for
the quantum computation. Therefore, most systems that are thought
of as possible realizations of a quantum computer can also be used
to build quantum chains. Of course there has to be some coupling between
the qubits. This is typically easy to achieve in solid state systems,
such as Josephson junctions with charge qubits~\cite{Bruder2005a,Romito2,Tsomokos2006},
flux qubits~\cite{Bruder,Bruder2005}\index{flux qubits} (see also
Fig.~\ref{fig:A-quantum-chain}) or quantum dots dots using the electrons~\cite{Loss1998,Greentree2004}
or excitons~\cite{DAmico,Lovett} as qubits. Other systems where quantum chain
Hamiltonians can at least be \emph{simulated} are NMR qubits~\cite{Suter2006,Jones,Zhang2005,Zhang2007}
and optical lattices~\cite{Garcia-Ripoll2003}. Such a simulation
is particularly useful in the latter case, where local control is
extremely difficult. Finally, qubits in cavities~\cite{Falci2005,Paternostro2005}
and coupled arrays of cavities were considered \cite{Bose,Plenioa,Bose2007}.
For more fundamental questions, such as studies of entanglement
transfer, non-locality and coherent transport, the quantum chains
could also be realized by systems which are not typically thought
of as qubits, but which are \emph{natural spin chains.} These can
be molecular systems~\cite{EXCITON} or quasi-1D solid state materials~\cite{Motoyama1996,Gambardella2002}.

\subsection{Excitation conserving chains and correlation functions\index{transfer functions}}

Most investigations on quantum state transfer with quantum chains
up to date are concentrating on Hamiltonians that conserve the number
of excitations (some more general cases are considered in \cite{Plenio,Plenio2006,Bruder2005,Kay}).
These Hamiltonians are \emph{much} easier to handle both analytically
and numerically, and it is also easier to get an intuition of the
dynamics. Furthermore, they occur quite naturally as a coupling between
qubits in the relevant systems. We stress though that there is \emph{no
fundamental} reason to restrict quantum chain communication to this
case. We denote the eigenstates of the Pauli matrix $\sigma_{z}$
as $\sigma_{z}|0\rangle=|0\rangle$ and $\sigma_{z}|1\rangle=-|1\rangle.$
The conservation of excitations means that $\sum_{k}\sigma_{z}^{(k)}$
commutes with the Hamiltonian $H$ of the system, where $\sigma_{z}^{(k)}$
is the Pauli matrix acting on the $k$th qubit of the chain. The state
$|\boldsymbol{0}\rangle\equiv|0\rangle^{(1)}\otimes|0\rangle^{(2)}\otimes\cdots\otimes|0\rangle^{(N)}$
with zero excitations is always an eigenstate of $H.$ We assume that
the system is initially in this state, which could be the case if
it is the ground state (as in the ferromagnetic case or in the presence
of a strong field in $z$ direction), or which could be achieved \emph{actively}
using an appropriate cooling protocol~\cite{YUASA1,Burgarth2007,Burgarth2006}.
The original state transfer protocol then proceeds as follows:

\begin{enumerate}
\item The state of the first qubit of the chain is swapped with an arbitrary
and unknown state $\rho.$ This is assumed to happen
much faster than the time-scale of the interaction of the
chain. It has recently been shown~\cite{Bruder} that this is not
a fundamental restriction, and that finite switching times can even slightly
improve the fidelity if they are carefully included in the protocol.
But this requires to solve the full time-dependent Schr\"odinger equation,
and introduces further parameters to the model (i.e. the raise and
fall time of the couplings). For the sake of simplicity, we will therefore
stick to the original assumption.
\item After time evolution $U=\exp\left(-iHt\right)$ for the time $t$
the quantum state at the end $N$ of the chain is picked up. Again this
is assumed to happen very fast.
\item If the chain is used multiple times, it has to be brought back to
the initial state $|\mathbf{0}\rangle$ in order to avoid memory effects~\cite{Werner2005}.
\end{enumerate}
The quality of the transfer is quantified by the fidelity between
the input state $\rho$ and the state that is picked up at the receiving
end. Since it still depends on the state $\rho$ that is being sent,
one can either average over all input states, or consider its minimum
over all input states. We are interested in the latter, which is easily
seen~\cite{Burgarth2006} to be \begin{align}
F_{0} & (t)=|\langle\sigma_{-}^{(N)}(t)\sigma_{+}^{(1)}\rangle|^{2},\label{eq:transfer}\end{align}
where $\sigma_{-}^{(N)}(t)=U^{\dag}(t)\sigma_{-}^{(N)}U(t)$ and the
expectation value is taken in the state $|\mathbf{0}\rangle.$ Eq.~(\ref{eq:transfer})
provides an interesting and intuitive connection between the time
dependent two-point correlation functions and state transfer for excitation
conserving systems. In statistical mechanics, one is typically interested
in the thermodynamic limit of these functions and in their scaling
for long time and distance. In the context of state transfer, their
short time behavior and the height and width of their maxima are the
most interesting quantities. Typically, due to the dispersion on the
chain, the fidelity of these devices is not good enough to make them
feasible~\cite{Bose2003,Linden2004,Landahl2005}.

\subsection{Advanced schemes}

Shortly after the initial proposal~\cite{Bose2003} it has been shown
that there are ways to achieve even \emph{perfect state transfe}r
with arbitrarily long chains. 
The Heisenberg model chosen by Bose
features many typical aspects of coherent transport, i.e. the wave-like
behavior, the dispersion, and the almost-periodicity of the fidelity.
These features do not depend so much on the specific choices of the
parameters of the chain, such as the couplings strengths. There are however \emph{specific couplings} for quantum chains that
show a quite different time evolution, and it was suggested in~\cite{Landahl2004}
and independently in~\cite{Lambropoulos2004a} to use these to achieve
a \emph{perfect} state transfer. These values for engineered couplings
also appear in a different context in~\cite{Cook1979,Peres1985}.
The time evolution features an additional \emph{mirror symmetry:}
the wave-packet disperses initially, but the dispersion is reversed
after its center has passed the middle of the chain. This approach
has been extended by various authors~\cite{Eisert2004,Kay,Landahl2005,Ericsson2005,Bose2005,Yung2006,Sun2006a,Sun2005c,Sun2005a,Lambropoulos2006,Lambropoulos2004,Kay2006,Kay2006a,Stolze2005,Jing-Fu2006,Ekert2004,Kostak2007},
and many other choices of parameters for perfect or near perfect state
transfer in various settings were found~\cite{Kay2006,Stolze2005,Ekert2004}.
A different approach of tuning the Hamiltonian was suggested in~\cite{Semiao2005,Bednarska,Bednarska2005}.
There, only the first and the last couplings $j$ of the chain are
engineered to be \emph{much weaker} than the remaining couplings $J$
of the chain, which can be quite arbitrary. The fidelity can be made
arbitrarily high by making the edge coupling strengths weaker. Some specific types of quantum chains
which show high fidelity for similar reasons were also investigated~\cite{Sun2005,Bose2006,Bose2006a,Pasquale2006,Venuti2007,Roncaglia2006}.
The disadvantage of reliying on specific couplings strengths is that they the fidelity is typically not stable against disorder~\cite{Romito2,Mont,Chin}.

A second approach is to encode the information in multiple qubits.
It was suggested first in~\cite{Linden2004}. There, it was assumed
that the chain consists of three sections: one part
controlled by the sending party, one ''free'' part 
and one part controlled by the receiving
party. The sender encodes the qubit not only in a single qubit of
the chain, but in a \emph{Gaussian-modulated superposition} of his
qubits. These Gaussian packets are known to have minimal dispersion.
Likewise, the receiver performs a decoding operation on all qubits
he controls. Near-perfect fidelity can be reached. Other strategies
that involve multi-qubit encodings or decodings were suggested in~\cite{DUALRAIL,RANDOMRAIL,MULTIRAIL,MEMORYSWAP}

Finally, a number of authors found ways of improving the fidelity
by time-dependent control of some parameters of the Hamiltonian. In~\cite{Haselgrove2005}
it is shown that if the end couplings can be controlled as arbitrary
(in general complex valued) smooth functions of time the encoding
scheme~\cite{Linden2004} could be \emph{simulated} without the requirement
of additional operations and qubits. Another possibility to achieve
perfect state transfer is to have an Ising interaction with additionally
pulsed global rotations~\cite{Jones,Fitzsimons2006,Raussendorf2005}.
Further related methods of manipulating the transfer by global fields
were reported in \cite{Monteiro2006,Boness2006,Sun2006,Maruyama,Korepin2005,Yang2006,Eckert2007}.

\section{Optimal spin chain communication in the presence of coupling fluctuations}

\subsection{Setup}

In~\cite{Burgarth} it was shown that that the fidelity can be improved
easily by applying in certain time-intervals two-qubit unitary gates
at the receiving end of the chain. These gates act as \emph{valves}
which take probability amplitude out of the system without ever putting
it back. We remark that this seeming time-irreversible behavior does
not contradict quantum mechanics, because the gates depend on the
step of the protocol. The required sequence is determined \emph{a
priori} by the Hamiltonian of the system and by the time-intervals
of the protocol. Arbitrarily high fidelity is guaranteed by a convergence
theorem for all non-Ising coupling types that conserve the number
of excitations. This includes disordered systems, though the time-scale
of the convergence will be very large if there is localization~\cite{Burrell2007}.
For short chains or weak disorder however Anderson localization does
not play any role and our protocol performs very stable - opposed
to most other schemes. An experimental demonstration of this iterative approach
has recently been given in a NMR system~\cite{Zhang2007}.%
\begin{figure}[htbp]
\begin{centering}\includegraphics[width=0.7\columnwidth]{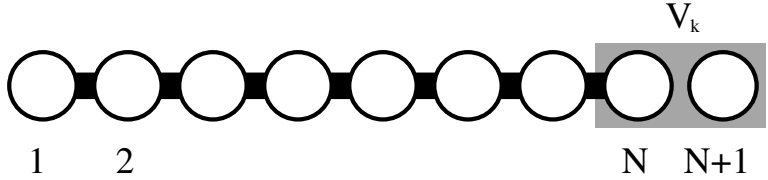}\par\end{centering}

\caption{\label{fig:setup}A quantum chain (qubits $1,2,\cdots,N$) and a
target qubit ($N+1$). By applying a sequence of two-qubit unitary
gates $V_{k}$ on the last qubit of the chain and the target qubit,
a state transfer of arbitrarily high fidelity can be achieved between the $1$st and $N+1$th qubit.}
\end{figure}

We now very briefly describe the valve method from a practical point
of view, i.e. how to determine the required operations from the Hamiltonian
of the system. The proofs and details are described elsewhere~\cite{Burgarth}.
Notice first that we consider $N+1$ qubits: the first $N$ are part
of the chain, and the $N+1$ one is the target qubit to which we transfer.
This qubit is not part of the chain and assumed to be \emph{uncoupled}. The
valve operations take place between the last qubit of the chain and
the target qubit (see Fig.~\ref{fig:setup}). We define the states
$|\mathbf{n}\rangle\equiv\sigma_{+}^{(n)}|\mathbf{0}\rangle$ with
exactly one excitation in the qubit $n.$ The time interval between
the $k$th step and the $k-1$th step of the protocol is $t_{k}$
with the corresponding time evolution operator of the chain $U_{k}=\exp\left(-iHt_{k}\right).$
Finally we define the unnormalized states $|\phi_{k}\rangle$ recursively by $|\phi_{k+1}\rangle=PU_{k}|\phi_{k}\rangle$
with ${|\phi}_{0}\rangle=|\mathbf{1}\rangle$ and $P=1-|\mathbf{N}\rangle\langle\mathbf{N}|.$
The valve gates are two-qubit gates acting on qubits $N$ and $N+1$. Their ideal sequence is expressed in the canonical basis
$\{|00\rangle,|01\rangle,|10\rangle,|11\rangle\}$ by 
\begin{equation}
\label{gates}
V_{k}=(F_0^k)^{-1/2}\left(\begin{array}{cccc}
1 &  0 & 0 & 0\\
0 & (F_0^{k-1})^{-1/2} & \left\langle \mathbf{N}|U_k|\phi_{k-1}\right\rangle ^{*}& 0 \\
0  & -\left\langle \mathbf{N}|U_k|\phi_{k-1}\right\rangle & (F_0^{k-1})^{-1/2} & 0\\
0  & 0 & 0 & 1\end{array}\right),\end{equation}
where the minimal fidelity for quantum state transfer at the $k$th step
of the protocol is given by\begin{equation}
F_{0}^{k}=1-\langle\phi_{k}|\phi_{k}\rangle.\end{equation}
This is a generalization of the time-dependent two-point correlation function
Eq.~(\ref{eq:transfer}) in the presence of active gating. 
Ideally the gates act in a time-scale much shorter than the evolution
of the chain, though this can be generalized~\cite{Bruder,Burgarth}.
For a given Hamiltonian one can easily compute the required
$V_{k}$ as a function of the time interval $t_{k}.$ The required
optimization may take various constraints into account: low number of
gates, high fidelity, short total time, regular intervals and stability
with respect to timing fluctuations.

\subsection{Fluctuations}

Now we consider the following: suppose one has computed a good sequence
of gates $V_{k}$ for the Hamiltonian $H$ that one planned to implement.
This could be an engineered Hamiltonian even in order to have a high
initial fidelity. When the chain is built, its real Hamiltonian will
be different from $H$ due to the finite precision of the implementation
parameters (such as the distance of the qubits, electric or magnetic
fields, and so on). This time-independent error can in principle be
compensated by performing a simple type of state tomography~\cite{RANDOMRAIL}
and by re-adjusting the gates $V_{k}$ to the new Hamiltonian. But
what happens if there are time-dependent errors,  caused by voltage fluctuations
or small displacements of the qubits? Alternatively, what happens
if -say in mass production- one is ignorant about even the static
errors, and refuses to perform tomography and a re-adjusting of the
gates for each individual chain? For these questions it is clearly
interesting to study what happens if the sequence $V_{k}$ is applied
to the \emph{wrong} Hamiltonian. In principle, this could be a arbitrary time-dependent
Hamiltonian. For simplicity we restrict ourselves to slowly varying Hamiltonians
which are approximately constant during \emph{one} application of the protocol,
but which have changed when the chain is used the next time. This assumption
allows us to solve the problem for time-independent Hamiltonians and then average
over different realizations.
The gates $V_{k}$ now no longer function as valves: they also put
probability amplitude from the target qubit back into the chain. Since
they are defined recursively, it could be that the error is amplified
during the protocol, and that the fidelity as a function of the number
of steps $k$ will strongly decrease after some critical value.

\begin{figure}[ht]
\begin{centering}\includegraphics[width=0.5\paperwidth]{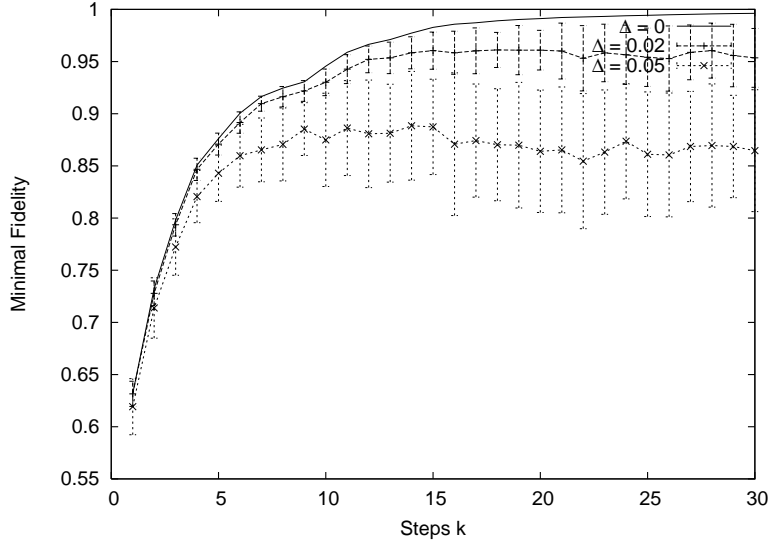}\par\end{centering}

\caption{\label{fig:fidelity2}Minimal fidelity $F_{0}^{k}$ for a $XX$ chain
of length $N=20$ as a function of the steps $k$ of the protocol.
Shown is the perfect case ($\Delta=0$) and two perturbed systems
with different fluctuation strength ($\Delta=0.02,0.05$). In the
latter cases, we give the average and standard deviation for $100$
samples.}
\end{figure}

\begin{figure}[ht]
\begin{centering}\includegraphics[width=0.5\paperwidth]{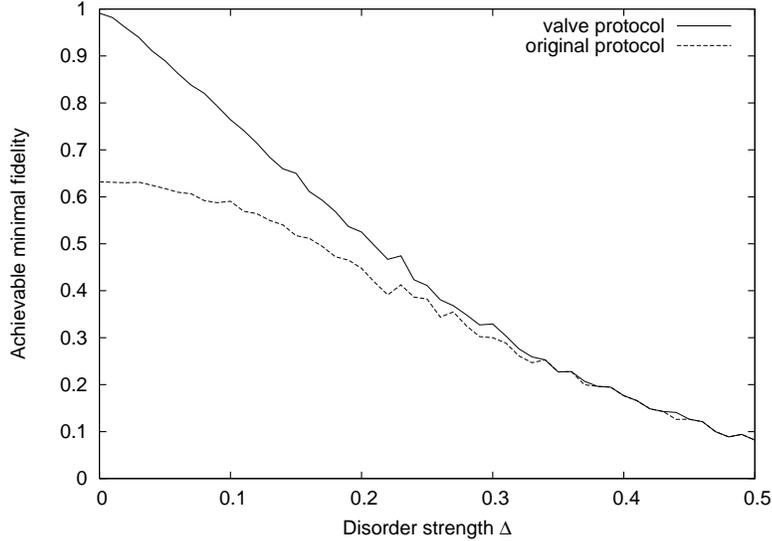}\par\end{centering}

\caption{\label{fig:fidelity3}Fidelity that can be achieved using maximally $20$ steps
of the valve protocol for a $XX$ chain of length $N=20$, i.e. $\mbox{max}_{k<20}F_{0}^{k}$. 
Shown is the case where the ideal sequence is applied to \emph{unknown} disorder as a function of the error strength
$\Delta$ (averaged over $100$ samples). As a comparison, we also
give the fidelity $F_{0}$ of the original protocol~\cite{Bose2003} as a function of unknown disorder.
Up to $\Delta\approx0.3$ the fidelity can be improved using the valve. }
\end{figure}

For the numerical analysis, we have considered a $XX$ Hamiltonian \begin{equation}
\label{han}
H(\underline{\delta})=\sum_{n=1}^{N-1}(1+\delta_{n})\left[\sigma_{x}^{(n)}\sigma_{x}^{(n+1)}+\sigma_{y}^{(n)}\sigma_{y}^{(n+1)}\right]\end{equation}
with coupling fluctuations $\underline{\delta}=(\delta_{1},\ldots,\delta_{N-1}).$
A good sequence of gates $V_{k}$ is determined numerically for the
ideal case $\underline{\delta}=\underline{0}.$ Then the protocol with these $V_k$ is
applied to a \emph{perturbed} Hamiltonian where the $\delta_{n}$ are uniformly
distributed in the interval $[-\Delta,\Delta],$ and the resulting minimal
fidelity is computed. This is repeated over $100$ samples and averaged.
Rather than a break-down of the fidelity after a critical number of steps, we
observe a saturation: at some point, the gain of applying another
gate is equal to the loss that is caused by the error (Fig.~\ref{fig:fidelity2}).
A similar behavior was already observed in the case of a switchable
interaction~\cite{Burgarth}. It is then interesting to compute
the maximum of $F_{0}^{k}$ with respect to $k$
and its dependence on the strength of the error. The results
shown in Fig.~\ref{fig:fidelity3} are very motivating: the decrease
is almost linearly in $\Delta$ over a wide range. Moreover, our protocol is capable of
improving the {}``natural'' performance of the chain (given by Eq.~(\ref{eq:transfer})) for fluctuations up to $\Delta=0.3.$%
 We obtained similar results also for Gaussian distributed disorder and for on-site disorder.

\section{Conclusion}
We have shown numerically that the valve protocol is surprisingly robust with respect
to known and \emph{unknown} disorder (such as slowly variying time-dependent disorder). This is a 
strong advantage with respect to the dual-rail scheme~\cite{RANDOMRAIL} which is only
capable of dealing with known disorder. It remains open to study in more detail time-dependent
Hamiltonians and to take into account realistic gating times, though we do not expect
a qualitative difference to the results discussed here.

\begin{acknowledgement}

I would like to thank the organizers of the 383. Wilhelm und Else
Heraeus Seminar for inviting me to contribute to this stimulating
workshop. I am very grateful to V. Giovannetti and S. Bose for the
long-standing collaboration. The work is finacially supported by the Swiss
National Science Foundation (SNSF).

\end{acknowledgement}

\end{document}